\documentclass[%
reprint,
superscriptaddress,
 amsmath,
 amssymb,
 aps,
 prl
]{revtex4-1}
\usepackage{graphicx}
\usepackage{subfig}
\usepackage{dcolumn}
\usepackage{bm}
\usepackage{caption}
\usepackage{balance}
\usepackage{bbm}
\usepackage{hyperref}
\hypersetup{
    colorlinks=true,
    linkcolor=red,
    filecolor=red,
    urlcolor=red,
    citecolor=blue, 
}

\usepackage[font=small,labelfont=bf]{caption}
\usepackage[symbol]{footmisc}
\usepackage{stackengine}

\newcommand{\bra}[1]{\left\langle #1 \right|}
\newcommand{\ket}[1]{\left|#1\right\rangle}

\newcommand{\braz}[1]{\langle #1 |}
\newcommand{\ketz}[1]{|#1\rangle}

\makeatletter
\newcommand*{\rom}[1]{\expandafter\@slowromancap\romannumeral #1@}
\makeatother

\begin{document}


\title{Signs of the rates in the Lindblad master equations can always be arbitrarily determined}

\author{\small Le Hu}
\email{lhu9@ur.rochester.edu}
\affiliation{Department of Physics and Astronomy, University of Rochester, Rochester, New York 14627, USA}
\affiliation{Institute for Quantum Studies, Chapman University, 1 University Drive, Orange, CA 92866, USA}
\author{\small and Andrew N. Jordan}

\affiliation{Institute for Quantum Studies, Chapman University, 1 University Drive, Orange, CA 92866, USA}
\affiliation{Department of Physics and Astronomy, University of Rochester, Rochester, New York 14627, USA}
\date{\today}

\begin{abstract}
Determining the Markovianity and non-Markovianity of a quantum process is a critical problem in the theory of open quantum systems, as their behaviors differ significantly in terms of complexity. It is well recognized that a quantum process is Markovian if and only if the quantum master equation can be written in the standard Lindblad form with all rates nonnegative for all time. However, here we present a striking result that \textit{any} finite-dimensional open quantum system dynamics can be described by a quantum master equation in the Lindblad form with all rates nonnegative for all time. In fact, it can be shown that one can arbitrarily decide the sign of the rates in any case at any time interval. Note that here we take an unconventional approach where the quantum master equation we construct will in general be state-dependent, which means that the Hamiltonian, jump operators and rates will all depend on the current state of the density matrix $\rho(t)$. Our findings raise serious questions on the current criterion in determining Markovianity and non-Markovianity in open quantum system dynamics.
\end{abstract}
\maketitle


\textbf{Introduction.}---Quantum systems that interact with their environment are called open quantum systems. Since no quantum systems are completely isolated, all realistic quantum systems are open, which means that their evolution is generally non-unitary. As a result, one cannot describe their evolution by the Schr\"{o}dinger equation alone, but has to rely on other means such as quantum master equations. It can be shown \cite{de2017dynamics} that all quantum master equations can be cast into the standard Lindblad form,
\begin{equation} \label{eq1}
	\dot{\rho}=-i[H, \rho]+\sum_{i=1}^{d^{2}-1} \gamma_{i}\left(L_{i} \rho L_{i}^{\dagger}-\frac{1}{2}\left\{L_{i}^{\dagger} L_{i}, \rho\right\}\right),
\end{equation}
where the Hamiltonian $H$, jump operators $L_i$ and rates $\gamma_i$ are potentially time-dependent. An important problem in the theory of open quantum systems is determining the Markovianity of a Lindblad master equation, since Markovian systems are usually significantly easier to describe. It is well recognized \cite{de2017dynamics, PhysRevA.89.042120, RevModPhys.88.021002, shrikant2023quantum, PhysRevLett.103.210401, PhysRevLett.105.050403, pang2017abrupt, groszkowski2023simple} that a quantum master equation in the standard Lindblad form is Markovian if and only if $\gamma_i(t) \geq 0$ for all time $t$, corresponding to the complete positive and trace-preserving (CPTP) process \cite{gorini1976completely, lindblad1976generators, RevModPhys.88.021002}.

However, in this Letter, we are going to show a striking result, that \textit{any} finite, $d$-dimensional continuous open quantum system dynamics can be described by the quantum master equation in the standard Lindblad form with nonnegative rates $\gamma_i(t) \geq 0$  for all $i$ and all time. Moreover, we can prove exactly the same conclusion, but this time with nonpositive rate $\gamma_i(t) \leq 0$ for all $i$ and all time. Actually, it turns out that one can even decide freely which $\gamma_i(t)$ is going to be nonpositive, and which is going to be nonnegative, in any case. This finding raises serious questions on the current criterion in determining Markovianity. 

In the following, we are going to prove the above statement. 

\textbf{Proof.}---Here we will focus on the 4-$d$ case for the purpose of better illustration, but the same reasoning can be easily carried over to arbitrarily finite dimensional systems, demonstrated later. To start with, let us first consider the simplified case that the Hamiltonian $H(t)=0$ and density matrix $\rho_D(t)$ is diagonalized, such that $\rho_D= \text{diag}(p_1, p_2, \dots, p_d)$ and $\dot{\rho}_D=\text{diag}(f_1, f_2, \dots, f_d)$ (in our case, $d=4$).
We define the following jump operators, 
\begin{equation}
\begin{aligned}
a_1&=\begin{pmatrix}
	0&0&0&0\\
	1&0&0&0\\
	0&0&0&0\\
	0&0&0&0
\end{pmatrix};a_2=\begin{pmatrix}
	0&0&0&0\\
	0&0&0&0\\
	0&1&0&0\\
	0&0&0&0
\end{pmatrix};a_3=\begin{pmatrix}
	0&0&0&0\\
	0&0&0&0\\
	0&0&0&0\\
	0&0&1&0
\end{pmatrix}\\
a_4&=\begin{pmatrix}
	0&0&0&0\\
	0&0&0&0\\
	1&0&0&0\\
	0&0&0&0
\end{pmatrix}; a_5=\begin{pmatrix}
	0&0&0&0\\
	0&0&0&0\\
	0&0&0&0\\
	0&1&0&0
\end{pmatrix}; a_6=\begin{pmatrix}
	0&0&0&0\\
	0&0&0&0\\
	0&0&0&0\\
	1&0&0&0
\end{pmatrix},
\end{aligned}
\end{equation}
and their conjugate transpose $a^\dagger_i$'s. We also denote the quantum channel $\Phi_{a_i}$, associated with $a_i$, and quantum channel $\Phi_{a^\dagger_i}$, associated with $a^\dagger_i$, as 
\begin{equation}
\begin{aligned}
	\Phi_{a_{i}}[\rho(t)]&=\gamma_{i}(t)\left(a_{i} \rho(t) a_{i}^{\dagger}-\frac{1}{2}\left\{a_{i}^{\dagger} a_{i}, \rho(t)\right\}\right)\\
	\Phi_{a^\dagger_{i}}[\rho(t)]&=\gamma^\prime_{i}(t)\left(a^\dagger_{i} \rho(t) a_{i}-\frac{1}{2}\left\{a_{i} a^\dagger_{i}, \rho(t)\right\}\right),
	\end{aligned}
\end{equation}
Then one can calculate $\Phi_{a_i}$'s and $\Phi_{a^\dagger_i}$'s. For instance,
\begin{widetext}
\begin{equation}\label{eq4}
\begin{aligned}
\Phi_{a_1}[\rho_D(t)]&=\begin{pmatrix}
	-\gamma_1 p_1 &0&0&0\\
	0&\gamma_1 p_1&0&0&\\
	0&0&0&0\\
	0&0&0&0
\end{pmatrix}; \quad \Phi_{a^\dagger_1}[\rho_D(t)]=\begin{pmatrix}
	\gamma^\prime_1 p_2 &0&0&0\\
	0&-\gamma^\prime_1 p_2&0&0&\\
	0&0&0&0\\
	0&0&0&0
\end{pmatrix}\\
\Phi_{a_2}[\rho_D(t)]&=\begin{pmatrix}
	0 &0&0&0\\
	0&-\gamma_2 p_2&0&0&\\
	0&0&\gamma_2 p_2&0\\
	0&0&0&0
\end{pmatrix}; \quad \Phi_{a^\dagger_2}[\rho_D(t)]=\begin{pmatrix}
	0 &0&0&0\\
	0&\gamma^\prime_2 p_3&0&0&\\
	0&0&-\gamma^\prime_2 p_3&0\\
	0&0&0&0
\end{pmatrix}.\\
\end{aligned}
\end{equation}
\end{widetext}
There are a few critical properties of the matrices $\Phi$'s presented above:
\begin{enumerate}
	\item Each $\Phi$ is diagonalized, and generally contains exactly two non-zero entries;
	\item Each $\Phi$ is traceless, meaning that if one of the non-zero entry equals $+g$ then the other one must equal $-g$.
	\item $\Phi_{a_i}$ and $\Phi_{a^\dagger_i}$ are linearly dependent with each other for a given $i$, but they have the opposite sign if $\gamma_i$ and $\gamma^\prime_i$ have the same signs.
	\item $\Phi_{a_i}$'s are generally linearly independent with each other for different $i$. Similarly for $\Phi_{a^\dagger_i}$'s.
\end{enumerate}
Note that the above properties also apply to the $d$-dimensional case, where we will have $\sum_{k=1}^{d-1} k=\frac{d(d-1)}{2}$ linearly independent $a_i$'s, and $\frac{d(d-1)}{2}$ linearly independent $a^\dagger_i$'s, resulting in a total number of $d(d-1)$ jump operators. On the other hand, to select two eigenvalues out of $d$, there are ${d \choose 2}=\frac{d(d-1)}{2}$ different ways. Since there are exactly $\frac{d(d-1)}{2}$ $\Phi_{a_i}$'s (or $\Phi_{a^\dagger_i}$'s), the linearly independency of $\Phi_{a_i}$'s (or $\Psi_{a^\dagger_i}$'s) means that $\Phi_{a_i}$'s (or $\Psi_{a^\dagger_i}$'s) span the space formed by all possible ``2-eigenvalue channel'', i.e. the quantum process that exactly two eigenvalues of the density matrix changes.

Recall that we need to solve the Eq.\,(\ref{eq1}) for nonnegative (later generalized to other cases) $\gamma_i$'s and $\gamma^\prime_i$'s, assuming arbitrary physical and known $p_i$'s and $f_i$'s, via the jump operators $a_i$'s and $a^\dagger_i$'s defined above. To do so, we need to solve 
$\dot{\rho}_{D}=\sum_{i}\left(\Phi_{a_{i}}[\rho_{D}(t)\right]+\Phi_{a_{i}^{\dagger}}\left[\rho_{D}(t)\right])$, or equivalently,
\begin{equation}
	f_i=F_i(\gamma_1, \gamma_2, \dots, \gamma_{d(d-1)/2}, \gamma^\prime_1, \gamma^\prime_2, \dots, \gamma^\prime_{d(d-1)/2})
\end{equation}
where $F_i$, a function of $\gamma$'s and $\gamma^\prime$'s, is the $i$th diagonal element of the master equation, whose explicit expression can be easily calculated. All off-diagonal elements of both sides vanish. Note that this is an inhomogeneous undetermined linear system of equations, which means that if there exists a solution, then there are infinitely many solutions. 

While there are some algebraic ways (e.g. row reduction) to solve the system of equations, here we take another approach. We divide $f_i$'s into two categories: the nonnegative ones, $f_{+,1}, f_{+,2}, \dots,f_{+,j} \geq 0$, and negative ones, $f_{-,1}, f_{-,2}, \dots,f_{-,k} < 0$, such that $j+k=d$ (in our case, $d=4$). Note that if all $f_i$'s are zero, then the evolution is unitary and all $\gamma$'s vanish. To proceed, let us first focus on $f_{-,1}$. Since $f_{-,1}$ is negative, its corresponding eigenvalue $p_{-,1}$ of $\rho(t)$ decreases. Such decreased amount of eigenvalue must be compensated by the same amount elsewhere by $f_{+}$'s so as to ensure $\operatorname{Tr}(\dot{\rho}_D)=0$. The key idea here is that we have the freedom to decide which $f_+$ is going to get the compensation by what amount, and each different choice corresponds to a different solution of $\gamma$'s.

Let us say $f_{+,1}$ is going to get the compensation as much as $f_{-,1}$ can afford and $f_{+,1}$ can receive. By the aforementioned analysis, we know that there always exists an $\Phi_{a_l}$ or $\Phi_{a^\dagger_l}$ for some $l$ such that the $\Phi_{a_l}$ or $\Phi_{a^\dagger_l}$ can deliver the compensation with nonnegative rate $\tilde{\gamma}_1 \equiv \gamma_l$ or $\tilde{\gamma}_1 \equiv\gamma^\prime_l$. For instance, in $d=4$ case, if $f_2$ is negative, and $f_1$ is positive and will be compensated, then by Eq.\,(\ref{eq4}) we should choose $\Phi_{a^\dagger_1}$ (but not $\Phi_{a_1}$); if $f_3$ is positive and will be compensated, then we should choose $\Phi_{a_2}$ (but not $\Phi_{a^\dagger_2}$). In general, there are three possible outcomes of the compensation:
\begin{enumerate}
	\item $f_{+,1}$ is exactly compensated by $f_{-,1}$, which means $|f_{+,1}|= |f_{-,1}|$. In this case, both $f_{+,1}$ and $f_{-,1}$ are ruled out for future compensations.
	\item $f_{+,1}$ is under-compensated, which means $|f_{+,1}|> |f_{-,1}|$. In this case, $f_{-,1}$ is ruled out for future compensations, whereas $f_{+,1}$ will still enjoy future compensations, and its value is updated to $f_{+,1}+f_{-,1}$.
	\item $f_{+,1}$ is over-compensated, which means $|f_{+,1}|< |f_{-,1}|$. In this case, $f_{+,1}$ is ruled out for future compensations, whereas $f_{-,1}$ will still enjoy future compensations, and its value is updated to $f_{+,1}+f_{-,1}$.
\end{enumerate}
As can be seen, at the end of a single round of compensation, at least one of the $f$'s will be ruled out for the future compensations. We proceed the same process for the rest pairs of $f_+$'s and $f_-$'s, e.g. $f_{+,1}$ and $f_{-,2}$, or $f_{-,1}$ and $f_{+,2}$, etc. In each round, we will solve for a nonnegative rate $\tilde{\gamma}_i$, and rule out at least one of the $f$'s. In the last round, the $f_{+}$ and $f_{-}$ will exactly cancel out each other, so two $f$'s are guaranteed to be eliminated. Since there are in total of $d$ number of $f$'s, as a result, we will have up to $d-1$ rounds of compensation, in which we will solve for nonnegative rates $\tilde{\gamma}_1, \tilde{\gamma}_2, \dots, \tilde{\gamma}_{d-1}$. This also implies that as few as $d-1$ jump operators are sufficient to describe any finite, $d$-dimensional open quantum system dynamics. This finding echos our recent work \cite{hu2023probabilistic}, in which we concluded that as few as $d-1$ unitary jump operators are sufficient to describe any $d$-dimensional open quantum system dynamics.

Moreover, we can show by the same method that \textit{any} quantum master equation can be written in the standard Lindblad form such that all rate $\tilde{\gamma}_i(t) \leq 0$ for all time. Actually, we can even choose which $\tilde{\gamma}_i$'s are going to be nonnegative, and which are going to be nonpositive, freely, in any case at any time interval. This is because for a designated sign of $\tilde{\gamma}_i$, we can always choose from $\Phi_{a_j}$ and $\Phi_{a^\dagger_j}$ for some $j$ and at least one of them will deliver the compensation since $\Phi_{a_j}$  and $\Phi_{a^\dagger_j}$ has the opposite sign. 

After finishing all compensation processes, we will have a list of rates $\tilde{\gamma}_1, \tilde{\gamma}_2, \dots, \tilde{\gamma}_{d-1}$, which are potentially time-dependent, and corresponding jump operators $\tilde{a}_1, \tilde{a}_2, \dots, \tilde{a}_{d-1}$, so that we can write down the master equation for the diagonalized density matrix,
\begin{equation}\label{eq6}
\dot{\rho}_D=\sum_{i=1}^{d-1}\tilde{\gamma}_i \left( \tilde{a}_i \rho_D \tilde{a}_i^\dagger-\frac{1}{2}\{\tilde{a}_i^\dagger \tilde{a}_i, \rho_D\}\right).
\end{equation}
To obtain the master equation for the non-diagonal case, we use the same trick we developed in \cite{hu2023probabilistic}, by establishing an explicit correspondence between $\rho_D$ and $\rho$, and $\dot{\rho}_D$ and $\dot{\rho}$, 
\begin{equation}
	\rho_{D}(t)=\mathbb{U}_{t}^{\dagger} \rho(t) \mathbb{U}_{t}
\end{equation}
\begin{equation}
	\dot{\rho}_{D}(t)=\mathbb{U}_{t}^{\dagger}(\dot{\rho}(t)+i[H(t), \rho(t)]) \mathbb{U}_{t},
\end{equation}
and plugging in $\dot{\rho}_D$ and $\rho_D$ into Eq.\,(\ref{eq6}). The $\mathbb{U}_t$ denotes the unitary matrix diagonalizing $\rho(t)$, and $H(t)$ denotes the Hamiltonian that can solve all instantaneous eigenvectors $\ket{\psi_i(t)}$ of $\rho(t)$, i.e. $H(t) \ket{\psi_i(t)}=i \ket{\partial_t \psi_i(t)}$ for $i=1,2, \dots, d$. An explicit form of $H(t)$, which is state-dependent, can be given by \cite{hu2023quantum},
\begin{equation}
H(t)=i\sum_{i=1}^d \ketz{\partial_t \tilde{\psi}_i(t)}\braz{\tilde{\psi}_i(t)},
\end{equation}
where  $|\tilde{\psi}_{i}(t) \rangle \equiv e^{i \phi_{i}(t)} |\psi_{i}(t) \rangle$ and $\phi_i(t)=\int-i\left\langle\partial_{t} \psi_{i}(t)|\psi_{i}(t)\right\rangle d t$.
The above Hamiltonian, which can be proven to be Hermitian by taking the time derivative of both sides of $\mathbbm{1}=\sum_{i=1}^d\ketz{\tilde{\psi}_i(t)}\braz{\tilde{\psi}_i(t)}$, is optimal in the sense that it has the minimum Hilbert-Schmit norm \cite{hu2023probabilistic} $\|H(t) \|_{HS}=\operatorname{Tr}(H^2(t))$. One can also take $H(t)=i\sum_{i=1}^d \ketz{\partial_t {\psi}_i(t)}\braz{{\psi}_i(t)}$ if such optimization is unnecessary. 

After plugging in $\dot{\rho}_D$ and $\rho_D$ into Eq.\,(\ref{eq6}), we can obtain the master equation, 
\begin{equation} \label{eq10}
	\dot{\rho}=-i[H(t), \rho(t)]+\sum_{i=1}^{d-1} \tilde{\gamma}_{i}\left(A_{i, t} \rho A_{i, t}^{\dagger}-\frac{1}{2}\left\{A_{i, t}^{\dagger} A_{i, t}, \rho\right\}\right),
\end{equation}
where $A_{i,t}=\mathbb{U}_{t} \tilde{a}_{i} \mathbb{U}_{t}^{\dagger}$. The master equation obtained in this way, which can describe \textit{any} continuous, $d$-dimensional open quantum system dynamics, has only linearly, $d-1$ many $\tilde{\gamma}_i$ terms, and can have arbitrary designated sign of $\tilde{\gamma}_i$, including the case that all $\tilde{\gamma}_i$'s are nonnegative or nonpositive. Note that different from usual master equations, the master equation Eq.\,(\ref{eq10}) is generally state-dependent, which means that the $H(t), \tilde{\gamma}_i$ and $A_{i,t}$ all depend on the current state of the density matrix $\rho(t)$. 

\textbf{Generalization to $d$-dimensional case.}---To show that the proof applies to arbitrary $d$-dimensional case, we denote $a_{ij} = \ket{i}\bra{j}$, $a^\dagger_{ij}=\ket{j}\bra{i}$, with $j<i\leq d$, and $\rho_D=\operatorname{diag}(p_1, p_2, \dots, p_d)$. Then we have,
\begin{equation}
\begin{aligned}
	\Phi_{a_{ij}}[\rho_D]&= p_j \lambda_{ij}(\ket{i}\bra{i}-\ket{j}\bra{j})\\
	\Phi_{a^\dagger_{ij}}[\rho_D]&= -p_i \lambda^\prime_{ij}(\ket{i}\bra{i}-\ket{j}\bra{j}),
\end{aligned}
\end{equation}
where we have used,
\begin{equation}
\begin{aligned}
	a_{ij}\rho_D=p_j a_{ij}&, \quad \rho_Da_{ij}=p_i a_{ij},\\
	a^\dagger_{ij} \rho_D=p_i a^\dagger_{ij}&, \quad  \rho_Da^\dagger_{ij}=p_ja^\dagger_{ij}, \\
	a_{ij}a^\dagger_{ij}=\ket{i}\bra{i}&, \quad a^\dagger_{ij}a_{ij}=\ket{j}\bra{j},\\
	\ket{i}\bra{i}\rho_D=p_i \ket{i}\bra{i}&, \quad \rho_D \ket{i}\bra{i}=p_i \ket{i}\bra{i}.
\end{aligned}
\end{equation}
From the expressions of $\Phi_{a_{ij}}$ and $\Phi_{a^\dagger_{ij}}$, it is immediately clear that if we want to describe a ``2-eigenvalue channel'', where the $i$th eigenvalue changes by some amount and $j$th eigenvalue changes by the negative of that amount, there always exists an $\Phi$ with which we can describe the process. In this particular case, we should either choose $\Phi_{a_{ij}}$ or $\Phi_{a^\dagger_{ij}}$ for the description, depending on whether we want our $\lambda$ to be negative or positive. The rests of the arguments follow exactly the same as illustrated earlier.

\textbf{Example - Jaynes-Cummings model.} Here we demonstrate an example in $d=2$ where all $\gamma$'s are made nonnegative and nonpositive, respectively, for all time. We consider the Jaynes-Cummings model under the rotating wave approximation, which describes the Rabi oscillation of a two-level atom in an cavity, 
\begin{equation}
	H_{S E}=\hbar \omega_{c} a^{\dagger} a+\hbar \omega_{a} \frac{\sigma_{z}}{2}+\frac{\hbar \Omega}{2}\left(a \sigma_{+}+a^{\dagger} \sigma_{-}\right),
\end{equation}
where $\sigma_\pm=\sigma_x \pm i \sigma_y$. For simplicity, we take $\omega_c=\omega_a=\omega$ and $\hbar=\Omega=1$. The model, which is conventionally considered as highly non-Markovian, can be solved exactly \cite{berman2011principles}, and the reduced density matrix $\rho_S(t)$ of the atom can be found by
\begin{equation}
	\rho_{S}(t)=\left(\begin{array}{cc}\rho_{11}(0) \cos ^{2}\left(\frac{t}{2}\right) & \rho_{12}(0) \cos \left(\frac{ t}{2}\right) e^{-i \omega t} \\ \rho_{21}(0) \cos \left(\frac{t}{2}\right) e^{i \omega t} & 1-\rho _{11}(0) \cos^{2}\left(\frac{t}{2}\right)\end{array}\right),
\end{equation}
where we have assumed that the cavity is initially in the vacuum state.
Alternatively, one can describe the above dynamics by the following master equation, 
\begin{equation}
\begin{aligned}
	\dot{\rho}=-i[H(t), \rho]&+\gamma_{1}(t)\left(\tilde{\sigma}_{-} \rho \tilde{\sigma}_{+}-\frac{1}{2}\left\{\tilde{\sigma}_{+} \tilde{\sigma}_{-}, \rho\right\}\right)\\
	&+\gamma_{2}(t)\left(\tilde{\sigma}_{+} \rho \tilde{\sigma}_{-}-\frac{1}{2}\left\{\tilde{\sigma}_{-} \tilde{\sigma}_{+}, \rho\right\}\right),
	\end{aligned}
\end{equation}
where $H(t)=\sum_{i=1}^{2}\ket{\partial_t \psi_i(t)}\bra{\psi_i(t)}$, $\tilde{\sigma}_\pm = \mathbb{U}_t \sigma_{\pm}\mathbb{U}^\dagger_t$, and $\ket{\psi_i(t)}$ is the instantaneous eigenvector of $\rho(t)$. Let $\rho_D(t)$ which is diagonalized by $\mathbb{U}_t$ be given by $\rho_D(t)=\operatorname{diag}(\lambda_1(t), \lambda_2(t))$. If we want both $\gamma$'s to be nonnegative, we can take
$\gamma_1(t)=\max (-\frac{\dot{\lambda}_1}{\lambda_1}, 0)$ and $\gamma_2(t)=\max (\frac{\dot{\lambda}_1}{1-\lambda_1},0)$; if we want both $\gamma$'s to be nonpositive, we can take $\gamma_1(t)=\min(-\frac{\dot{\lambda}_1}{\lambda_1}, 0)$ and $\gamma_2(t)=\min(\frac{\dot{\lambda}_1}{1-\lambda_1},0)$. If initially, $\rho_S(t)$ is already diagonalized (i.e. $\rho_{12}(0)=\rho_{21}(0)=0$), then the solution have a simple explicit form. In this case, $\mathbb{U}_t=\mathbb{U}^\dagger_t=\mathbbm{1}$ such that $\tilde{\sigma}_\pm = \sigma_\pm$, and we can take $H=\frac{1}{2}\omega \sigma_z$. If we want both $\gamma$'s to be nonnegative, we can take
\begin{equation}\label{eq16}
\begin{aligned}
	\gamma_1(t)&=\left\{\begin{array}{ll} \tan\frac{t}{2} \geq0, & \text {  } t\in[0+2n \pi,(2n+1) \pi), n\in \mathbb{N}  \\ 0 & \text { } \text{otherwise}\end{array}\right.\\
	\gamma_2(t)&=\left\{\begin{array}{ll} \alpha(t) \geq 0, & \text {  } t\in[(2n+1) \pi,(2n+2) \pi), n\in \mathbb{N}  \\ 0 & \text { } \text{otherwise,}\end{array}\right.
\end{aligned}
\end{equation}
and if we want both $\gamma$'s to be nonpositive, we can take
\begin{equation}
\begin{aligned}
	\gamma_1(t)&=\left\{\begin{array}{ll} \tan\frac{t}{2} \leq0, & \text {  } t\in[(2n+1) \pi,(2n+2) \pi), n\in \mathbb{N}  \\ 0 & \text { } \text{otherwise}\end{array}\right.\\
	\gamma_2(t)&=\left\{\begin{array}{ll} \alpha(t) \leq0, & \text {  } t\in[0+2n \pi,(2n+1) \pi), n\in \mathbb{N}  \\ 0 & \text { } \text{otherwise,}\end{array}\right.
\end{aligned}
\end{equation}
where $\alpha(t)=\frac{\rho_{11}(0)\sin t}{\rho_{11}(0)\cos t+\rho_{11}(0)-2}.$
Intuitively, Eq.\,(\ref{eq16}) can be understood as a qubit being in contact with a cold bath during the time interval $t \in[0+2 n \pi,(2 n+1) \pi), n \in \mathbb{N}$, with rate $\gamma_1(t)$, and with a hot bath during the other time, with rate $\gamma_2(t)$. Both processes are conventionally thought to be Markovian. Note that $\gamma$ can be singular by approaching infinity. The reason is that the eigenvalue $\lambda_1(t)$, which could be zero at certain moment, appears in the denominator. We stress that such a singular value, while annoying, will not hamper the description of the dynamics. In fact, such singularity has been reported in the literature before \cite{hu2023probabilistic, PhysRevLett.104.070406, PhysRevLett.103.210401}.

\textbf{Conclusions.}---Since for any given quantum master equation, we can always rewrite it in the Lindblad form given by Eq.\,(\ref{eq10}) and choose whatever the sign of $\gamma_i$ we want, the current notion of Markovianity and non-Markovianity breaks down, at least in the sense of the signs of $\gamma_i$'s. Moreover, the current interpretation \cite{RevModPhys.88.021002} which associates non-Markovianity with the information backflow from the environment to the system also becomes questionable. As such, a future reexamination of the those notions becomes necessary and essential.

\textbf{Acknowledgement.}---We are grateful to Shengshi Pang for valuable discussions. This work was supported by the Army Research Office (ARO) under Grant No. W911NF-22-1-0258.


\appendix
\bibliographystyle{ieeetr}
\bibliography{apssamp3}

\begin{thebibliography}{10}

\bibitem{de2017dynamics}
I.~De~Vega and D.~Alonso, ``Dynamics of non-markovian open quantum systems,'' {\em Reviews of Modern Physics}, vol.~89, no.~1, p.~015001, 2017.

\bibitem{PhysRevA.89.042120}
M.~J.~W. Hall, J.~D. Cresser, L.~Li, and E.~Andersson, ``Canonical form of master equations and characterization of non-markovianity,'' {\em Phys. Rev. A}, vol.~89, p.~042120, Apr 2014.

\bibitem{RevModPhys.88.021002}
H.-P. Breuer, E.-M. Laine, J.~Piilo, and B.~Vacchini, ``Colloquium: Non-markovian dynamics in open quantum systems,'' {\em Rev. Mod. Phys.}, vol.~88, p.~021002, Apr 2016.

\bibitem{shrikant2023quantum}
U.~Shrikant and P.~Mandayam, ``Quantum non-markovianity: Overview and recent developments,'' {\em Frontiers in Quantum Science and Technology}, vol.~2, p.~1134583, 2023.

\bibitem{PhysRevLett.103.210401}
H.-P. Breuer, E.-M. Laine, and J.~Piilo, ``Measure for the degree of non-markovian behavior of quantum processes in open systems,'' {\em Phys. Rev. Lett.}, vol.~103, p.~210401, Nov 2009.

\bibitem{PhysRevLett.105.050403}
A.~Rivas, S.~F. Huelga, and M.~B. Plenio, ``Entanglement and non-markovianity of quantum evolutions,'' {\em Phys. Rev. Lett.}, vol.~105, p.~050403, Jul 2010.

\bibitem{pang2017abrupt}
S.~Pang, T.~A. Brun, and A.~N. Jordan, ``Abrupt transitions between markovian and non-markovian dynamics in open quantum systems,'' {\em arXiv preprint arXiv:1712.10109}, 2017.

\bibitem{groszkowski2023simple}
P.~Groszkowski, A.~Seif, J.~Koch, and A.~Clerk, ``Simple master equations for describing driven systems subject to classical non-markovian noise,'' {\em Quantum}, vol.~7, p.~972, 2023.

\bibitem{gorini1976completely}
V.~Gorini, A.~Kossakowski, and E.~C.~G. Sudarshan, ``Completely positive dynamical semigroups of n-level systems,'' {\em Journal of Mathematical Physics}, vol.~17, no.~5, pp.~821--825, 1976.

\bibitem{lindblad1976generators}
G.~Lindblad, ``On the generators of quantum dynamical semigroups,'' {\em Communications in Mathematical Physics}, vol.~48, pp.~119--130, 1976.

\bibitem{hu2023probabilistic}
L.~Hu and A.~N. Jordan, ``Probabilistic unitary formulation of open quantum system dynamics,'' {\em arXiv preprint arXiv:2307.05776}, 2023.

\bibitem{hu2023quantum}
L.~Hu and A.~N. Jordan, ``Quantum state driving along arbitrary trajectories,'' {\em Physical Review Research}, vol.~5, no.~3, p.~033045, 2023.

\bibitem{berman2011principles}
P.~R. Berman and V.~S. Malinovsky, {\em Principles of laser spectroscopy and quantum optics}.
\newblock Princeton University Press, 2011.

\bibitem{PhysRevLett.104.070406}
D.~Chru\ifmmode \acute{s}\else \'{s}\fi{}ci\ifmmode~\acute{n}\else \'{n}\fi{}ski and A.~Kossakowski, ``Non-markovian quantum dynamics: Local versus nonlocal,'' {\em Phys. Rev. Lett.}, vol.~104, p.~070406, Feb 2010.

\end{thebibliography}

\end{document}